\documentclass{tPHM2e}

\usepackage{epstopdf}
\usepackage{subfigure}

\usepackage{color}

\theoremstyle{plain}

\theoremstyle{definition}

\theoremstyle{remark}

\def\gsim{\mathop {\vtop {\ialign {##\crcr 
$\hfil \displaystyle {>}\hfil $\crcr \noalign {\kern1pt \nointerlineskip } 
$\,\sim$ \crcr \noalign {\kern1pt}}}}\limits}
\def\lsim{\mathop {\vtop {\ialign {##\crcr 
$\hfil \displaystyle {<}\hfil $\crcr \noalign {\kern1pt \nointerlineskip } 
$\,\,\sim$ \crcr \noalign {\kern1pt}}}}\limits}

\begin{document}



\title{Ubiquity of Unconventional Phenomena Associated with Critical Valence Fluctuations 
in Heavy Fermion Metals}

\author{
\name{K. Miyake\textsuperscript{a}$^{\ast}$\thanks{$^\ast$Corresponding author. Email: miyake@toyotariken.jp}
and S. Watanabe\textsuperscript{b}}
\affil{\textsuperscript{a}
Toyota Physical and Chemical Research Institute, Nagakute, Aichi 480-1192, Japan; \\
\textsuperscript{b}
Department of Basic Sciences, Kyushu Institute of Technology, Kitakyushu, 
Fukuoka 804-8550, Japan}
\received{December 16, 2016}
}

\maketitle

\begin{abstract}
Ubiquity of unconventional phenomena observed in a series of heavy fermion 
metals is discussed on the basis of an idea of critical valence fluctuations.  
After surveying experimental aspects of these unconventional behaviors in prototypical 
compounds, CeCu$_2$(Si,Ge)$_2$, under pressure, we propose that sharp valence crossover 
phenomena are realized in CeCu$_6$, CeRhIn$_5$, and Ce(Ir,Rh)Si$_3$ by tuning 
the pressure and the magnetic field simultaneously, on the basis of previous results 
for an extended Anderson lattice model with the Coulomb repulsion $U_{\rm fc}$ between 
localized f-electron and itinerant conduction electrons. 
\end{abstract}


\section{Introduction}
In the past decade or so, it gradually turned out that the critical-valence-transition or 
sharp-valence-crossover phenomena in heavy fermion metals is rather ubiquitous than 
first thought in the beginning of the present century. 
Indeed, unconventional quantum critical phenomena, which cannot be understood 
on the basis of the quantum criticality associated with magnetic transitions, have been observed in 
a series of heavy fermion metals, 
YbCu$_{5-x}$Al$_x$ (x=3.5) \cite{Bauer,Seuring}, 
YbRh$_2$Si$_2$ \cite{Trovarelli,Ishida}, YbAuCu$_4$ \cite{Yamamoto,Wada}, 
$\beta$-YbAlB$_4$ \cite{Nakatsuji}, 
$\alpha$-YbAl$_{1-x}$Fe$_x$B$_4$ (x=0.014 ) \cite{Kuga}, and quasi-crystal compound 
Yb$_{15}$Al$_{34}$Au$_{51}$ \cite{Deguchi} and quasi-crystal-approximant Yb$_{14}$Al$_{35}$Au$_{51}$ 
under pressure $P\simeq 1.8$GPa as well {\cite{Matsukawa}}. 
The non-Fermi liquid behaviors observed in these 
compounds can be explained in a coherent way by a scenario based on the critical valence 
fluctuations (CVF) using the mode-mode coupling approximation for CVF \cite{Watanabe1,Watanabe1a}, 
as summarized in Table\ \ref{Table:1}. 
{A} recent highlight is that the so-called $T/B$ scaling behavior of the magnetization, observed in 
$\beta$-YbAlB$_4$ \cite{Nakatsuji,Matsumoto,Matsumoto2}, and 
quasi-crystal related compounds Yb$_{15}$Al$_{34}$Au$_{51}$ {\cite{Deguchi2}}
and Yb$_{14}$Al$_{35}$Au$_{51}$ \cite{Matsukawa}, 
has been theoretically derived by taking into account 
the effect of the magnetic field in the mode-mode coupling approximation scheme  \cite{Watanabe2,Watanabe5}. 

\begin{table}
\caption{Theoretical results for a series of physical quantities by critical valence 
fluctuations (CVF) giving the exponent $\zeta$ as $0.5\lsim \zeta \lsim 0.7$ 
depending on the region temperature $T$ higher than $T_{0}$, the extremely small temperature 
scale (see \S6 and \S7), and unconventional criticality observed in a series of materials.  
{$T^{1.5}\to T$} in the column $\rho(T)$ means that $T$ dependence of $\rho(T)$ crosses over 
from $T^{1.5}$ at $T<T_{0}$ to $T$ at $T>T_{0}$ around $T\sim T_{0}$. 
The symbol * indicates that there is no available experiment. In the last low, theoretical 
prediction on the conventional criticality associated with antiferromagnetic (AF) quantum critical 
point (QCP) are shown, in which $-T^{1/2}$ and $-T^{1/4}$ indicate decrement from some constant 
values, respectively.\\
\ 
}
\label{Table:1}
\begin{center}
{\offinterlineskip
\halign{\strut\vrule#&\quad#\hfil\quad&
              \vrule#&\quad#\hfil\quad&
              \vrule#&\quad#\hfil\quad&
              \vrule#&\quad#\hfil\quad&
              \vrule#&\quad#\hfil\quad&
              \vrule#&\quad#\hfil\quad&
              \vrule#\cr
      \noalign{\hrule}
&Theories \&  Materials&& 
  $\rho(T)$ && $C(T)/T$ && $\chi(T)$&& $1/T_{1}T$&&
Refs. &\cr
      \noalign{\hrule}
& CVF && 
 {$T^{1.5}\to T$} && $-\log\,T$ && $T^{-\zeta}$ && $T^{-\zeta}$ &&
 \ \ \cite{Watanabe1,Watanabe1a,Watanabe4}&\cr
      \noalign{\hrule}
& YbCu$_{5-x}$Al$_{x}\,(x\simeq 1.5)$&& {$T^{1.5}\to T$} && 
 $-\log\,T$ && $T^{-0.66}$ && * &&
 \ \ \cite{Bauer,Seuring}&\cr
      \noalign{\hrule}
& YbRh$_2$Si$_2$ && 
  $T$ && $-\log\,T$ && $T^{-0.6}$ && $T^{-0.5}$&&
 \ \ \cite{Trovarelli,Ishida}&\cr
      \noalign{\hrule}
& $\beta$-YbAlB$_4$ &&  $T^{1.5}\to T$ &&
 $-\log\,T$ && $T^{-0.5}$ && * &&
 \ \ \cite{Nakatsuji} &\cr
      \noalign{\hrule}
& $\alpha$-YbAl$_{1-x}$Fe$_{x}$B$_4$\ ($x\simeq0.014$) &&  $T^{1.5}\to T$ &&
 $-\log\,T$ && $T^{-0.5}$ && * &&
 \ \ \cite{Kuga} &\cr
      \noalign{\hrule}      
& Yb$_{15}$Al$_{34}$Au$_{51}$ &&  
  $T$ && $-\log\,T$ && $T^{-0.51}$ && $T^{-0.51}$&&
 \ \ \cite{Deguchi}&\cr
      \noalign{\hrule}
& Yb$_{14}$Al$_{35}$Au$_{51}$ at $P\simeq 1.8$ GPa&&  
  $T$ && * && $T^{-0.51}$ && *&& 
 \ \ \cite{Deguchi2}&\cr
      \noalign{\hrule}
& AF QCP && 
  $T^{3/2}$ && $-T^{1/2}$ && $-T^{1/4}$ && $T^{-3/4}$ &&
 \ \ \cite{MoriyaTakimoto,Hatatani,Hatatani2}&\cr
      \noalign{\hrule}
}
}
\end{center}
\end{table}

The unconventional phenomena associated with sharp valence crossover have also been observed 
in a series of Ce-based heavy fermion metals{, 
while the complete systematic behaviors shown in Table\ \ref{Table:1} were not 
observed because such Ce-based compounds are not considered to be just at the valence criticality 
but in the sharp valence crossover region. 
CeCu$_2$Ge$_2$ is the first example in which}  
anomalous properties characteristic to the sharp valence crossover of Ce ion under pressure 
{around $P=P_{\rm v}$ was reported} \cite{Jaccard,Miyake0}. In particular, 
the superconducting transition temperature exhibits a drastic increase by about triple of that at 
around the magnetic critical pressure. 
{The $T$-linear dependence in the resistivity $\rho(T)-\rho_{0}$, with $\rho_{0}$ being the 
residual resistivity, was also ovserved at around $P=P_{\rm v}$, together with a huge enhancement 
of $\rho_{0}$.} 
After that, similar behaviors have been reported in CeCu$_2$Si$_2$ \cite{Holmes}, 
CeCu$_2$Si$_{1.8}$Ge$_{0.2}$ \cite{Yuan}, and CeRhIn$_5$ \cite{Knebel,Park}, 
which can be comprehensively understood on the basis of {a} valence crossover 
scenario \cite{Miyake1,Miyake2}.  It was also predicted \cite{Watanabe3,Watanabe3a} 
that the position of the critical point of valence transition is 
well controlled by the magnetic field, which opens a possibility of realizing the critical 
point by tuning both pressure and magnetic field simultaneously. Recently, a symptom of such a 
phenomenon was reported in CeCu$_6$ \cite{Hirose} which is considered to be located in the crossover 
region of valence transition \cite{Miyake1,Raymond}. This suggests that the puzzling 
non-Fermi liquid properties observed in CeCu$_{6-x}$Au$_x$ (x$\simeq$0.1) \cite{Loehneysen} 
should be revisited from the viewpoint of this CVF scenario.  

The essence of these intriguing phenomena cannot be captured within the so-called the 
Doniach paradigm \cite{Doniach} 
that is essentially based on the Kondo lattice picture in which the valence of 
Ce and Yb ion is fixed as C$^{3+}$ and Yb$^{3+}$.  In other words, we had to develop a conceptually 
new physics. The purpose of the present paper is to sketch the history and the present status of 
unconventional phenomena associated with sharp valence crossover or enhanced valence 
fluctuations in Ce-based heavy fermion metals, 
and to discuss how these phenomena are ubiquitous than thought previously. 
The organization of the paper is as follows:   

In Sect.\ 2, we review how the existence of critical 
point of valence transition in heavy fermion metals was recognized by showing the case of 
CeCu$_2$(Si,Ge)$_2$.  

In Sect.\ 3, we present some fundamental properties of the extended 
Anderson lattice model (including the Coulomb repulsion between f and conduction electrons)  
which is the minimal model for describing essential aspects of valence transition and 
valence fluctuations observed in Ce- or Yb-based heavy fermion metals.  

In Sect.\ 4, we show that the magnetic field gives appreciable influence on the valence transition 
of Ce ion discussing the case of CeCu$_6$.  

In Sect.\ 5, we discuss how the change of the Fermi surface observed in CeRhIn$_5$ under pressure 
is understood as a sharp valence crossover phenomenon in a unified fashion without relying on an idea of 
the destruction of c-f hybridization. 

In Sect.\ 6, summary is given.


\section{Sharp crossover in valence of Ce in CeCu$_2$(Si,Ge)$_2$}
In this section, we discuss a series of experimental evidence suggesting that the valence of Ce 
exhibits a sharp crossover under pressure in CeCu$_2$Si$_2$ \cite{Holmes}{,} 
{resulting in} a series of anomalous properties associated with the sharp valence crossover, 
while such a trend was first reported~\cite{Jaccard} and discussed theoretically~\cite{Miyake0} 
for CeCu$_2$Ge$_2$.  

\begin{figure}[h]
\begin{center}
\includegraphics[width=0.7\linewidth]{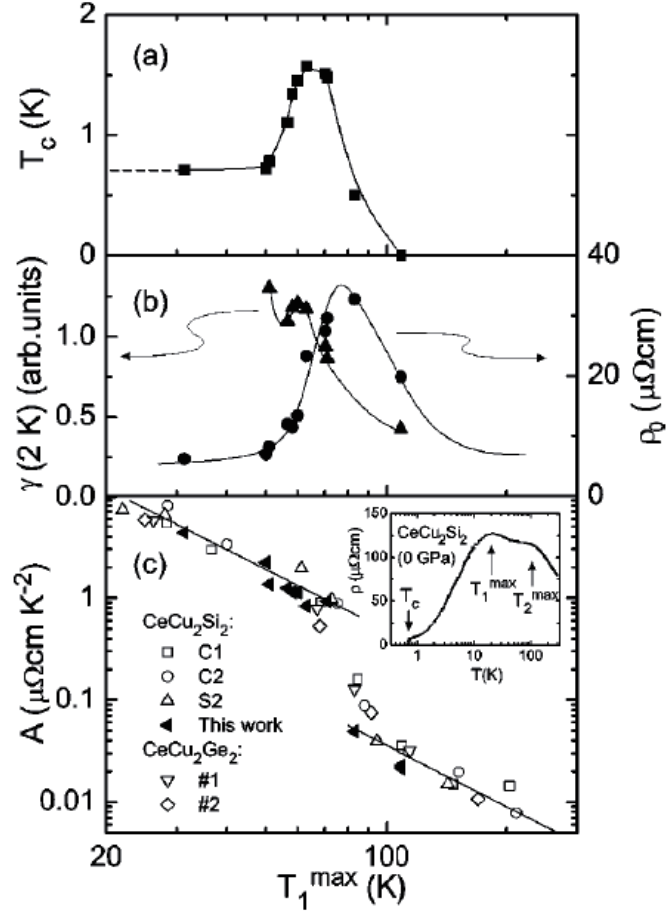}
\caption{
Horizontal axis $T_1^{\rm max}$ is the characteristic temperature at which the resistivity takes 
maximum, as shown in inset, and simulates the change in pressure $P$ which increases 
{toward} the right-hand side of the scale: 
(a) the bulk superconducting transition temperature, 
(b) the residual resistivity and $\gamma$ coefficient of the 
electronic specific heat, and
(c) the coefficient $A$ of the $T^2$-law of the resistivity.  The straight lines 
indicate that the expected $A\propto (T_1^{\rm max})^{-2}$ scaling is 
followed.  The maximum of $T_{\rm c}$ coincides with the start of 
the region where the scaling relation is broken, while the 
maximum in residual resistivity is situated in the middle of the collapse in $A$.  \cite{Holmes} 
}
\label{Fig:2}
\end{center}
\end{figure}

A drastic decrement in the $A$ coefficient of 
the $T^{2}$ term in the resistivity, $\rho(T)\simeq \rho_{0}+AT^{2}$,  by about two orders of 
magnitude around the pressure $P=P_{\rm v}\simeq4.5$ GPa, as shown in Fig.\  \ref{Fig:2}(c), 
suggests a sharp valence crossover occurs at around $P=P_{\rm v}$.   
Note that $T_{1}^{\rm max}$ is an increasing function of pressure $P$ and 
simulates the variation of $P$. 
The residual resistivity $\rho_{0}$ exhibits a sharp and pronounced peak there,  
as shown in Fig.\ \ref{Fig:2}(b).  
This implies that the effective mass $m^{*}$ of the quasiparticles also
decreases sharply there, since $A$ is scaled by $(m^{*})^{2}$ \cite{rf:KW}.   
This decrement of $m^{*}$ implies in turn a sharp change 
in the valence of Ce ion, deviating from Ce$^{3+}$, 
considering the fact that the following approximate (but canonical) 
formula holds in the strongly correlated limit \cite{RiceUeda,Shiba}: 
\begin{equation}
\label{eq:3m*nf}
{m^{*}\over m_{\rm band}} ={1-n_{\rm f}/2\over 1-n_{\rm f}},
\end{equation}
where $m_{\rm band}$ is the band mass without electron correlations, and 
$n_{\rm f}$ is the f-electron number per Ce ion.  

Such a sharp crossover in the valence of Ce ion  
gives rise to a sharp crossover of the so-called Kadowaki-Woods (KW) ratio \cite{rf:KW}, 
$A/\gamma^{2}$, where $\gamma$ is the Sommerfeld coefficient of
the electronic specific heat, from that of a strongly correlated class to a weakly correlated 
one.  Note that $\gamma^{-1}$ is related to the temperature $T_{1}^{\rm max}$,
where the resistivity $\rho(T)$ exhibits maximum, as shown in the inset of Fig.\ \ref{Fig:2}(c).  
This indicates, as discussed in Ref.\ \cite{rf:MMV}, that the mass enhancement due to 
the dynamical electron correlation is quickly lost at around $P=P_{\rm v}$ \cite{rf:MMV}.  

\begin{figure}
\begin{center}
\includegraphics[width=0.6\linewidth]{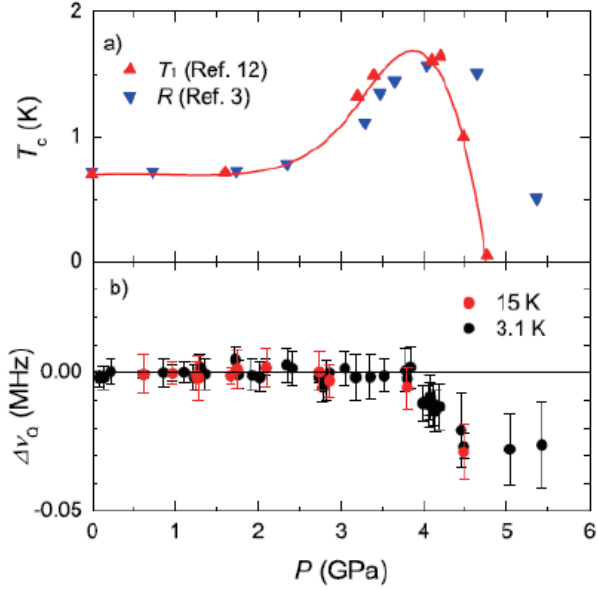}
\caption{(Color online) 
Pressure dependence of the superconducting transition temperature $T_{\rm c}$ and 
deviation from the linear $P$ dependence of background in the 
NQR frequency $^{63}\nu_{\rm Q}$ {in CeCu$_2$Si$_2$.~\cite{Kobayashi}}   
}
\label{Fig:Kobayashi}
\end{center}
\end{figure}

This physical picture based on the sharp valence crossover of Ce has been reinforced by 
$^{63}$Cu-NQR measurements  in CeCu$_2$Si$_2$ at temperature down to 
$T=3.1\,$K and under pressures up to $P=5.5\,$GPa covering 
$P_{\rm v}\simeq 4.5\,$GPa \cite{Fujiwara,Fujiwara2,Kobayashi}.  Indeed, the NQR frequency 
$^{63}\nu_{\rm Q}$ rather sharply deviates at above 4$\,$GPa from the linear $P$-dependence in the 
low pressure range ($P\le 3.5\,$GPa).  
The $P$-dependence of the deviation from the linear dependence in $^{63}\nu_{\rm Q}$ is 
shown in Fig.\ \ref{Fig:Kobayashi}b).   
This deviation was estimated to correspond to the change of the valence  
$\Delta n_{\rm f}={0.04}$ by the first principles calculations \cite{Kobayashi}.  
This degree of change is consistent with the decrease of mass enhancement by {$\sim3.9$
 ($\sim15.3$} in the $A$ coefficient of the resistivity{, as shown in Fig.\ \ref{Fig:2}(c)}), 
if the change in $n_{\rm f}$ would be from $n_{\rm f}=0.99$ to $n_{\rm f}={0.95}$.  

{
In Ref.\ \cite{Kobayashi}, the change of valence in Ce was estimated from that in NQR 
frequency $\nu_{\rm Q}$ as follows.  On the basis of the LDA calculations, the pressure ($P$) 
dependence of $\nu_{\rm Q}$ for ``LaCu$_{2}$Si$_{2}$'' and ``CeCu$_{2}$Si$_{2}$'' was 
estimated with the use of the observed $P$ dependence of the lattice parameter, 
where f electrons are assumed to be completely localized in the case of ``LaCu$_{2}$Si$_{2}$'', i.e., 
essentially in $(4{\rm f})^{0}$ configuration, and itinerant but not strongly correlated in 
the case of ``CeCu$_{2}$Si$_{2}$''. Both show a linear $P$ dependence with similar slopes: 
d$\nu_{\rm Q}/{\rm d}P\simeq0.103$MHz for ``LaCu$_{2}$Si$_{2}$'' and 
d$\nu_{\rm Q}/{\rm d}P\simeq0.089$MHz for ``CeCu$_{2}$Si$_{2}$''. However, observed NQR frequency 
$\nu_{\rm Q}$ of a real CeCu$_2$Si$_2$ is located in between because 4f electrons in 
CeCu$_2$Si$_2$ are not fully itinerant but have a localized character reflecting strong correlations 
among 4f electrons.  Indeed,  $\nu_{\rm Q}(P=0)\simeq 3.73$MHz for ``LaCu$_{2}$Si$_{2}$'' and 
$\nu_{\rm Q}(P=0)\simeq 3.02$MHz for ``CeCu$_{2}$Si$_{2}$'', while the observed one for the 
real CeCu$_{2}$Si$_{2}$ is  $\nu_{\rm Q}(P=0)\simeq 3.43$MHz.  
The deviation in $\nu_{\rm Q}(P)$ from a $P$ linear dependence shown in Fig.\ \ref{Fig:Kobayashi}(b) 
reflects a change of electronic contribution to $\nu_{\rm Q}$, while the linear $P$ dependence 
reflects the lattice contribution. The amount of deviation $\Delta\nu_{\rm Q}$ from $P=3.9$ to 
$P=4.5$ is about $\Delta\nu_{\rm Q}\simeq-0.03$MHz, which is $3.9$\% of the difference 
$[\nu_{\rm Q}($''CeCu$_{2}$Si$_{2}$''$)-\nu_{\rm Q}($''LaCu$_{2}$Si$_{2}$''$)]\simeq-0.77$MHz at around 
$P=3.9$-$4.5\,$GPa. The negative value of $\Delta\nu_{\rm Q}$ implies an increase in the 
itinerant character of 4f electrons, and the ratio of 4f electrons of acquiring itinerant character 
is estimated as $0.03/0.77\simeq0.04$ of the localized component, suggesting $\Delta n_{\rm f}\simeq 0.04$.  
}

The huge peak of  $\rho_{0}$ at $P\sim P_{\rm v}$, as shown in Fig.\ \ref{Fig:2}(b), 
can be explained  by the effect that the disturbance in the ratio of numbers of f and conduction 
electrons around impurity extends to the correlation length $\xi_{\rm v}$ which grows appreciably 
{toward} $P=P_{\rm v}$ giving rise to strong scattering of quasiparticles.   
The microscopic justification has been discussed in Ref.\ \cite{MM}{.}
This is in contrast to the effect of AF critical fluctuations on $\rho_{0}$ which is 
rather moderate, as discussed in Ref.\ \cite{MN}.  
Thus, the critical pressure $P_{\rm v}$ can be clearly manifested by 
the maximum of $\rho_0$.  

Subtle but systematic tendencies shown in Fig.\ \ref{Fig:2} near $P=P_{\rm v}$ are that the 
peak of the $T_{\rm c}$ and the Sommerfeld coefficient $\gamma(T=2~{\rm K})$ appears  
at slightly lower pressure than $P_{\rm v}$.  These behaviors can be understood on the basis of 
explicit theoretical calculations in which almost local valence fluctuations of Ce {are} 
shown to develop around the pressure where the sharp valence crossover occurs 
\cite{Holmes,Onishi1,Miyake1}{, as supported by the 
density-matrix-renormalization-group (DMRG) calculation \cite{WIM}.}    

\begin{figure}
 \begin{center}
 \includegraphics[width=0.7\columnwidth]{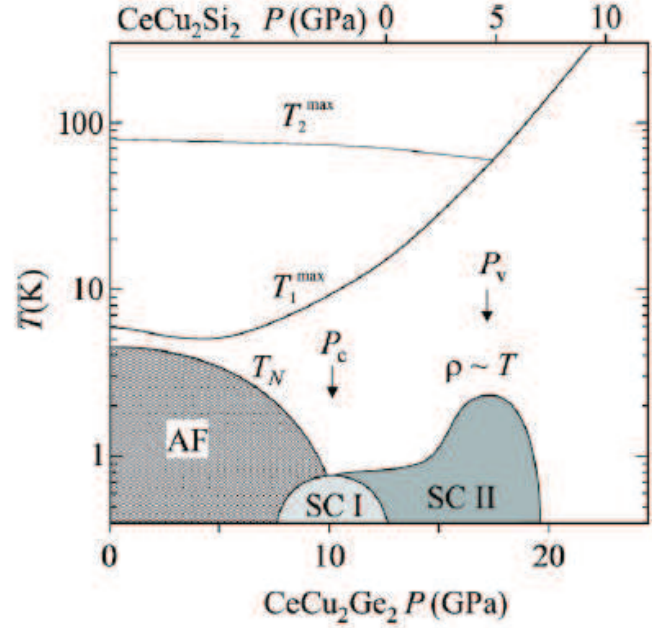}
 \caption{\label{Fig:PD_Holmes} 
Schematic $P$-$T$ phase diagram for CeCu$_2$(Si/Ge)$_2$ showing the two critical pressures 
 $P_{\rm c}$ and $P_{\rm v}$ \cite{Holmes}.  At $P_{\rm c}$, where the antiferromagnetic ordering 
temperature $T_N\rightarrow 0$, 
superconductivity in region SC\ I is mediated by 
 antiferromagnetic spin fluctuations; around $P_{\rm v}$, in the region SC\ II, 
valence  fluctuations provide the pairing mechanism and the resistivity is linear in temperature.  
The characteristic temperatures $T_{1}^{\rm max}$ and  $T_{2}^{\rm max}$ 
merge at a pressure $P\simeq P_{\rm v}$.
 }
 \end{center}
 \end{figure}

Another salient property associated with the sharp crossover of the valence is 
that the temperatures $T_{i}^{\rm max}$ ($i=1,2$), corresponding to two maxima in the 
the rsistivity $\rho(T)$, merge at $P\simeq P_{\rm v}$ as shown in Fig.\ \ref{Fig:PD_Holmes} 
\cite{Holmes}.  
Although the existence of the two peaks in $\rho(T)$ and {the} mergence 
{of them} under pressure can 
be shown theoretically as an effect of smearing crystalline-electric-field (CEF) splitting 
by the increase of the c-f hybridization under pressure in general \cite{Nishida}, 
it is non-trivial that the mergence occurs at around $P=P_{\rm v}$.  
This behavior was also observed in CeCu$_2$Ge$_2$ \cite{Jaccard}, CeAu$_2$Si$_2$ \cite{Jaccard2}, 
and CeAl$_2$ \cite{CeAl2}, suggesting a generic property associated with the sharp 
valence crossover of Ce ion as argued below \cite{Holmes}. 

A fundamental wisdom is that the so-called Kondo temperature $T_{\rm K}$, related 
to $T_{i}^{\rm max}$ ($i=1,2$), depends crucially on the 
degeneracy $(2\ell+1)$ of local f-state: $T_{\rm K}\sim D\exp[-1/(2\ell+1)N_{\rm F}|J|]$, 
where $D$ is {half} the bandwidth of conduction electrons, $N_{\rm F}$ the density of states of
conduction electrons at the Fermi level, and $J$ the c-f exchange coupling constant \cite{Yoshimori75}. 
Furthermore, even though the sixfold degeneracy of local 4f-state is lifted by the CEF effect, leaving 
the Kramers doublet ground state with the excited CEF levels with 
excitation energy $\Delta_{\rm CEF}$, the Kondo temperature 
$T_{\rm K}$ is still enhanced considerably by the effect of 
excited CEF level{s} \cite{Yamada84}. 

It is also crucial to take into account the fact that the practical degeneracy of CEF levels, 
relevant to the Kondo effect, is affected by the broadening $\Delta E$ of the 
lowest CEF level. If $\Delta E\ll\Delta_{\rm CEF}$, the degeneracy 
relevant to $T_{\rm K}$ is twofold.  On the other hand, 
if $\Delta E>\Delta_{\rm CEF}$, it increases to fourfold or sixfold.  
The level broadening is given by $\Delta E\simeq z\pi N_{\rm F}|V|^{2}$ 
where $|V|$ is the strength of c-f hybridization, and $z$ is the 
renormalization factor which gives the inverse of mass enhancement 
in the case of lattice system{s}.  Then, it is crucial that $\Delta E$ is 
very sensitive to the valence of Ce ion because $z^{-1}$ is essentially 
given by eq.\ (\ref{eq:3m*nf}). In particular, the factor $z$ 
increases from the tiny value in the Kondo regime, 
$z\sim(1-n_{\rm f})\ll 1$, and approaches to 1 in the so-called valence fluctuation regime.

Since the factor $\pi N_{\rm F}|V|^{2}\gg\Delta_{\rm CEF}$ in general for Ce-based heavy 
fermion metals, the ratio $\Delta E/\Delta_{\rm CEF}$, which is much smaller than 1 
in the Kondo regime, greatly exceeds 1 across the valence transformation around 
$P\sim P_{\rm v}$, {leading} to the increase of the practical  
degeneracy of f-state, {\it irrespective} of a sharpness of the 
valence transformation. Therefore, $T_{1}^{\rm max}$ should merge 
$T_{2}^{\rm max}$, which corresponds to fourfold or sixfold degeneracy of 
4 f-state due to the effect of  broadening of the CEF ground level.  
This explains why $T_{1}^{\rm max}$ 
increases and approaches $T_{2}^{\rm max}$ at around $P=P_{\rm v}$.

On the other hand, there have been trials of explaining this mergence of $T_{i}^{\rm max}$ ($i=1,2$) 
at around $P=P_{\rm v}$ as a phenomenon caused by a meta-orbital trans{ition} among 
CEF levels \cite{Hattori} or an interchange of CEF level scheme itself \cite{Pourovskii}.  
The latter theoretical prediction contradicts with experimental measurements both of 
the inelastic neutron scattering \cite{Horn} 
and the non-resonant X-ray scattering (NRXS) \cite{Willers,Rueff2} which shows that the ground state of 
the CEF levels in CeCu$_2$Si$_2$ is $\Gamma_{7}^{1}\simeq -0.88|\pm5/2\rangle+0.47|\mp3/2\rangle$.  
The former prediction \cite{Hattori} does not { immediately} contradict with the result of NRXS 
measurement \cite{Rueff2}, while it is not so evident whether the condition for the meta-orbital 
trans{ition} to occur is satisfied in the CEF states of CeCu$_2$Si$_2$. 
{Reference\ \cite{Hattori} assumes much larger c-f hybridization at 
the first-excited CEF level than that at the ground CEF level. However,} the CEF 
excited state with the c-f hybridization, larger than that of the ground CEF state, 
seems to correspond to the highest CEF level, 
$\Gamma_{7}^{2}\simeq 0.47|\pm5/2\rangle+0.88|\mp3/2\rangle$, 
with excitation energy about 360 K \cite{Horn}.

\section{Model for describing valence transition and fluctuations}
A canonical model for describing the valence transition due to electronic origin is an 
extended periodic Anderson model (EPAM) {that} take{s} into account 
the Coulomb repulsion $U_{\rm fc}$ between f and conduction electrons.  Explicit form of the 
EPAM is given as follows: 
\begin{eqnarray}
H_{\rm EPAM}
 &=& \sum_{{\bf k} \sigma}(\epsilon_{\bf k}-\mu) c_{{\bf k} \sigma}^{\dagger}c_{{\bf k} \sigma}
 +\varepsilon_{\rm f} \sum_{{\bf k} \sigma}f_{{\bf k} \sigma}^{\dagger}f_{{\bf k} \sigma} 
 +U_{\rm ff}\sum_i n_{i \uparrow}^{\rm f} n_{i \downarrow}^{\rm f}%
\nonumber \\
 & &\qquad+V\sum_{{\bf k} \sigma}(c_{{\bf k} \sigma}^{\dagger}f_{{\bf k} \sigma}^{}+{\rm h.c.})
 +U_{\rm fc}\sum_{i \sigma \sigma'}n_{i \sigma}^{\rm f}n_{i \sigma'}^{\rm c},
 \label{eq:1}
\end{eqnarray}
where $U_{\rm fc}$ {is} the f-c Coulomb repulsion, which turns out a main origin of valence transition 
or fluctuations.  The label $\sigma$ in Eq.\ (\ref{eq:1}) stands the degrees of freedoms 
of the Kramers doublet state of the ground CEF level.   This model has been discussed in 
a variety of context 
\cite{HewsonRiseborough, Schlottmann, CostiHewson, TakayamaSakai, Perakis, Khomskii}.   

The model without hybridization $V$ 
is called the Falicov-Kimball model (FKM) which ha{s} been a canonical model for discussing 
the valence state of rare earth ions \cite{FalicovKimball,Varma1}.  
The condition for the valence transition in the FKM is simply given by 
\begin{equation}
\varepsilon_{\rm f}+n_{\rm c}U_{\rm fc}=\mu,
\label{eq:2}
\end{equation}
where $\mu$ is the chemical potential or the Fermi level in the Kondo limit where f electrons 
are essentially singly occupied, and $n_{\rm c}$ is the number of conduction electrons at f-site. 
This relation expresses the competition between the energy level of f-electron modified by the mean field 
arising from  $U_{\rm fc}$, the f-c Coulomb repulsion, and the Fermi energy of conduction electrons.  

\begin{figure}[h]
\begin{center}
\rotatebox{0}{\includegraphics[width=0.7\linewidth]{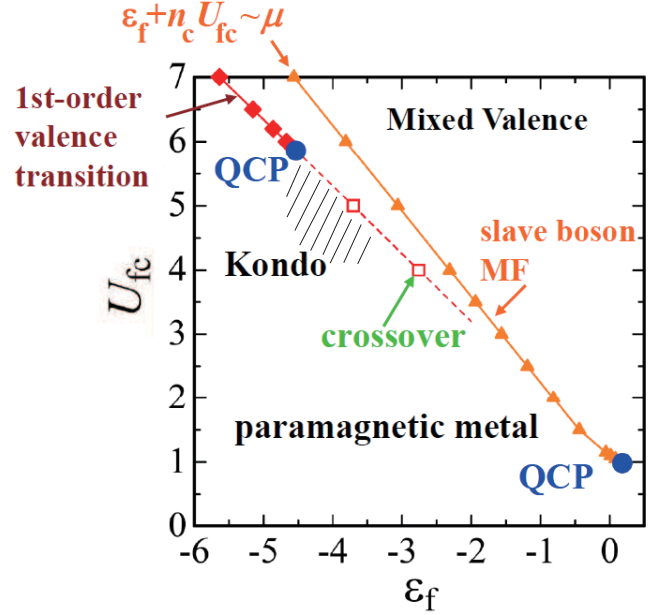}}
\caption{
Phase diagram at $T=0$ of the system described by the Hamiltonian [Eq.\ (\ref{eq:1})] 
in the $\varepsilon_{\rm f}$-$U_{\rm fc}$ plane. 
Triangles are the results by the slave-boson mean-field approximation \cite{Onishi1}, 
and diamonds and squares are 
those by DMRG calculations for one dimensional version of  the Hamiltonian [Eq.\ (\ref{eq:1})] 
in which $\epsilon_{k}=-2t\cos ka$ with $a$ being the lattice constant \cite{WIM}. 
Solid lines represent the first order valence transition, and dashed line 
represent{s} the valence crossover from  {the} Kondo to {mixed valence} regime.  Closed circles 
are critical end point of the first order valence transition{, i.e., quantum critical point (QCP) of 
valence transition}.  
In the shaded region, superconducting correlation with spin-singlet and inter-site paring dominates over 
both the SDW and CDW correlations in one dimensional model, 
suggesting that the d-wave superconductivity is stabilized in 
this region.  Parameters are $V/t=0.1$, 
$U_{\rm ff}/t=100$,  
and the total electron filling, ${\bar n}$, is fixed as ${\bar n}=7/8$, the same as 
used in Ref.\ \cite{Onishi1}. Unit of $U_{\rm fc}$ and $\varepsilon_{\rm f}$ is also $t$.  
}
\label{PD_DMRG}
\end{center}
\end{figure}

Although the existence of the hybridization $V$ makes it difficult to solve the problem, 
the condition [Eq.\ (\ref{eq:2})] remains to be valid in the mean-field level of approximation. 
Figure\ \ref{PD_DMRG} shows the ground state phase diagrams of the EPAM in 
the $\varepsilon_{\rm f}$-$U_{\rm fc}$ plane that {are} obtained  
by the mean-field approximation using the slave boson technique by taking into account 
the strong correlation effect ($U_{\rm ff}=\infty$) {\cite{Onishi1}}, and by the 
{DMRG} method for the one-dimensional version 
of the Hamiltonian Eq.\ (\ref{eq:1}) with the tight-binding dispersion for the conduction 
electrons, i.e., $\epsilon_{k}=-2t\cos ka$ with $a$ bing the lattice constant \cite{WIM}.  
The electron filling ${\bar n}\equiv (n_{\rm c}+n_{\rm f})/2$ is fixed as ${\bar n}=7/8$ in both calculations.  
A bunch of calculations on the Gutzwiller variational 
ansatz have also been performed \cite{Onishi2,Saiga, Sugibayashi,Kubo,Hagymasi}.  
The first order valence transition (FOVT) line is given essentially by the condition (\ref{eq:2}) 
{in the mean-field approximation}.  
The quantum critical end point (QCEP), i.e., the quantum critical point (QCP) of the FOVT 
shown by closed circles, shifts 
from the position given by the mean-field approximation to that given by asymptotically 
{\it exact} DMRG calculation 
{owing} to the strong quantum fluctuation effect.  In this approach based on EPAM 
{[Eq.\ (\ref{eq:1})]}, a series of electronic properties associated with the critical valence 
fluctuations can be {\it directly} calculated within a required accuracy as summarized in 
Table\ \ref{Table:1}.  

Although the DMRG calculation in one dimension {is} asymptotically exact, it cannot
give a definite conclusion about ordered state{s}. Nevertheless, we can obtain 
a useful information on the ordered 
state in higher dimensions by analyzing the dominant long-range correlation function.  As shown in 
Fig.\ \ref{PD_DMRG}, the superconducting tendency with spin-singlet and inter-site paring dominates 
over both the SDW and CDW tendencies in the Kondo regime, shaded region, near the crossover 
line.  This suggests that the d-wave superconductivity is stabilized in the Kondo regime near the 
crossover line, which is consistent with the result obtained by {the} theory 
on the basis of slave-boson mean-field solutions supplemented by Gaussian fluctuations around them 
in the case of three dimensional free dispersion for the conduction electrons, i.e., 
$\epsilon_{k}=k^{2}/2m$, and with the same electron filling ${\bar n}=7/8$ \cite{Onishi1}.

\section{Effect of magnetic field on valence transition and crossover: Case of CeCu$_6$}
In this section, the effect of the magnetic field on the critical valence transition or sharp crossover 
of valence in Ce-based heavy fermion metals is discussed. 
Since the valence transition or crossover is a phenomenon associated with a charge transfer process 
or sometimes it is referred to charge fluctuations phenomenon, 
it appears not to be affected considerably by the magnetic field. However, this is not the case. 
Indeed, the valence-transition temperature $T_{\rm v}\simeq 46$ K of  $\alpha$-$\gamma$ 
transition at ambient pressure in Ce$_{0.8}$La$_{0.1}$Th$_{0.1}$ (without magnetic field) 
is suppressed to $T_{\rm v}=0$\ K by the magnetic field of 
about 50 T \cite{Drymiotis}.  This fact suggests that the position of the QCEP of  {the} 
valence transition 
is {also} greatly influenced by the magnetic field, or the QCEP is easily induced by 
the magnetic field if the system is located near the QCEP.  Indeed, this is the case as discussed below.  

\begin{figure}
\begin{center}
\includegraphics[width=0.6\linewidth]{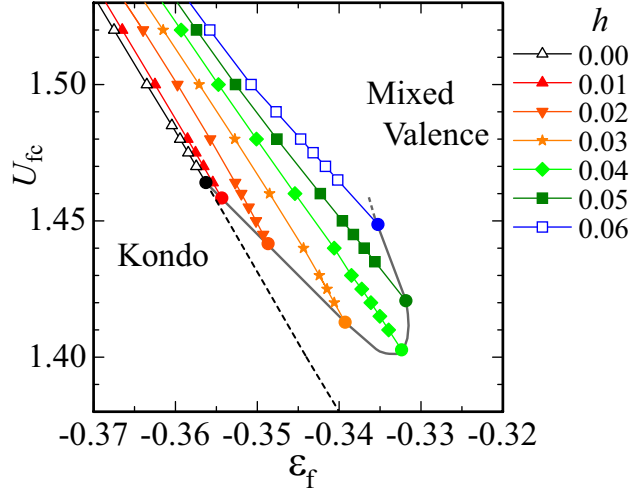}
\caption{\label{fig:hPD}(color online) Ground-state phase diagram in the plane of 
$U_{\rm fc}{-}\varepsilon_{\rm f}$ for $D=1$, $V=0.5$ at the electron filling ${\bar n}=7/8$. 
The FOVT line with a QCP for 
$h=0.00$ (open triangle), $h=0.01$ (filled triangle) $h=0.02$ (filled inverse triangle), 
$h=0.03$ (filled star), $h=0.04$ (filled diamond), 
$h=0.05$ (filled square) and $h=0.06$ (open square). 
The {curve} connects the QCP under $h$, which is a guide for the eyes. 
The dashed line represents the valence-crossover points at which 
$\chi_{\rm v}\equiv -(\partial n_{\rm f}/\partial \varepsilon_{\rm f})$ has a maximum 
as a function of $\varepsilon_{\rm f}$ for each $U_{\rm fc}$ at $h=0.00$ \cite{Watanabe3,Watanabe3a} {.}
}
\end{center}
\end{figure}

The effect of the magnetic filed $h$ applied along the $z$-direction is taken into account by 
introducing the Zeeman term to the EPAM Hamiltonian [Eq.\ (\ref{eq:1})] as 
\begin{equation}
H=H_{\rm EPAM}-h\sum_{i}(S_{{\rm f}i}^{z}+S_{{\rm c}i}^{z}), 
\label{eq:1_Zeeman}
\end{equation}
where $S_{{\rm f}i}^{z}$ and  $S_{{\rm c}i}^{z}$ are $z$-component of spin of f and conduction 
electrons, respectively, and the dispersion of conduction electrons in $H_{\rm EPAM}$ [Eq.\ (\ref{eq:1})] 
is set as $\epsilon_{k}=k^{2}/2m-D$, which extends from $-D$ to $D$.  
Figure\ \ref{fig:hPD} shows how the QCEP of valence transition in the ground s{t}ate moves 
by  {applying} magnetic field $h$ on the basis of slave boson 
mean{-field approximation} which properly takes 
into account the correlation effect due to the strong on-site Coulomb interaction $U_{\rm ff}$ 
in Eq.\ (\ref{eq:1_Zeeman}) in the ground state \cite{Watanabe3, Watanabe3a}.  
The unit of energy and magnetic field is taken as $D$, and the electron fillling is fixed as ${\bar n}=7/8$.  
The physical meaning of 
this sensitivity of the position of the QCEP is that the f-electron level of down spin, i.e., $S_{{\rm f}i}^{z}<0$, 
increases {toward} 
the Fermi level by the magnetic field, which promotes the valence transition, as discussed in 
Refs.\ \cite{Miyake1,Watanabe3, Watanabe3a}. 

\begin{figure}
\includegraphics[width=1.0\linewidth]{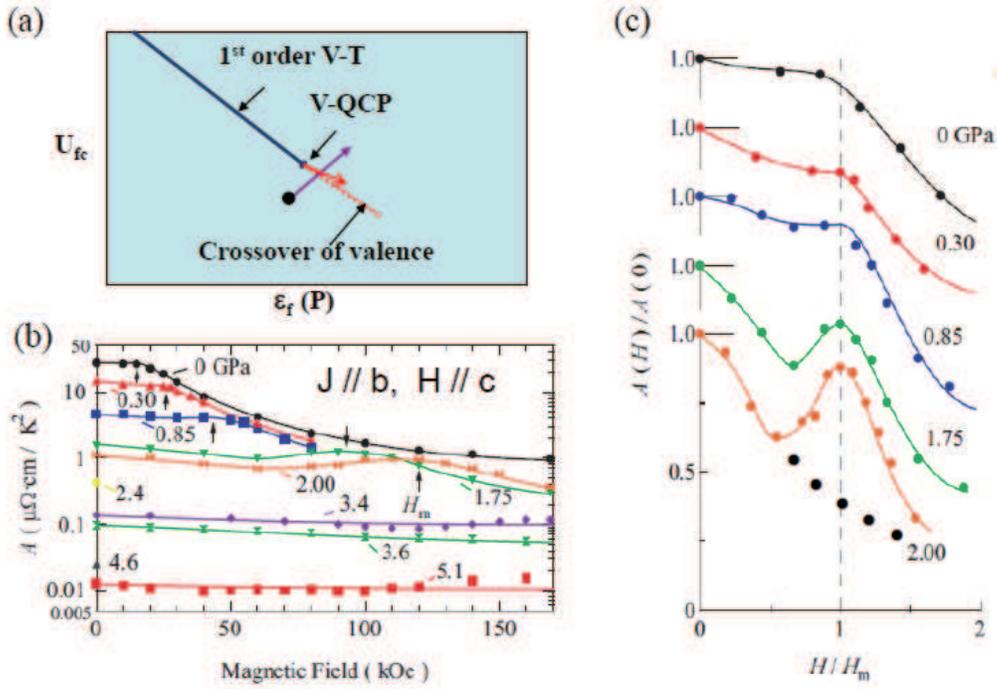}
\caption{\label{fig:PD_P_H}(color online) 
(a) Schematic view how the critical end point is tuned by 
pressure $P$ and magnetic field $H$. Large and small closed circles represent the position of the system 
and the QCEP, respectively.  A line with arrow from the system point indicates the locus of the system 
point as $P$ increases, and a {red} line with arrow from the QCEP indicates the locus of 
QCEP as $h$ increases. 
(b) The magnetic field ($H$) dependence of the $A$ coefficient of the $T^{2}$ term in the 
resistivity $\rho(T)$ in the logarithmic scale under a series of pressures $P$s \cite{Hirose}. 
{Vertical arrows indicate the metamagnetic field $H_{\rm m}$ for each pressure.}  
(c) Scaled relation between the $A$ coefficient and the magnetic field $H$ for a series of  
pressures $P$s \cite{Hirose}. 
A series of closed circles at $P=2.00$ GPa is a guide to the eye showing a possible background variation 
of $A$ not related to the valence crossover.  
 
} 
\end{figure}

This implies that the magnetic field is a good tuning 
parameter for searching the QCEP of valence transition. 
In particular, the QCEP can be {precisely} hit by simultaneously changing the magnetic field 
$h$ and the pressure $P$ if the system is located near the QCEP in the 
$\varepsilon_{\rm f}$-$U_{\rm fc}$ plane at the ambient state, as shown in Fig.\ \ref{fig:PD_P_H}(a). 
Indeed, since both $U_{\rm fc}$ and 
$\varepsilon_{\rm f}$ increase as $P$ increases in the Ce-based compounds, 
the point indicating the location of the system shifts {toward} right and upper direction 
as shown in Fig.\ \ref{fig:PD_P_H}(a).  Therefore, the locus of the QCEP due to applying $h$ 
intersect {s} with that due to applying $P$.  

Recently, a symptom of this phenomenon has 
been observed in CeCu$_6$ \cite{Hirose}. Figure \ref{fig:PD_P_H}(b) shows the magnetic field 
dependence of the $A$ coefficient of the $T^{2}$ term in the resistivity under a series of pressures $P$s. 
At ambient pressure, $A$ is a{n almost} monotonically decreasing function of $H$. 
By increasing pressure, it begins to show a {clear} maximum at $H_{\rm m}$ 
where the metamagnetic sharp increase in the magnetization is observed. 
Note that  $A$ is shown in a logarithmic scale. Variations of the $A$ coefficient is rather prominent 
if it is plotted in a linear scale, as shown in Fig.\ \ref{fig:PD_P_H}(c) where $A$ is scaled by $A(0)$ 
at ambient pressure and $H$ is scaled by the metamagnetic field $H_{\rm m}$. The peak structure 
becomes sharper and sharper as $P$ increases, which strongly suggests that $A$ diverges at the 
QCEP because the $T$ dependence of the resistivity should exhibit the $T$-linear dependence there.  
Experiments at higher pressure and magnetic field are highly desired.

Indeed, it has already been reported that CeCu$_6$ is located near the QCEP of valence transition 
and {a} sharp valence crossover occurs at $P\simeq 5$ GPa, 
although the decrement in the $A$ coefficient against the pressure is slightly moderate compared 
to {that in} CeCu$_2$Si$_2$ and CeCu$_2$Ge$_2$ \cite{Raymond,Miyake1}.
A characteristic temperature $T_{\rm F}^{*}$ (or the effective Fermi 
energy $\epsilon_{\rm F}^{*}$) of CeCu$_6$ is very low of the order of 10 K 
reflecting the heaviness of its effective mass.  This suggests that the QCEP is recovered by the 
magnetic field of the order of $T_{\rm F}^{*}$, i.e., $H\gsim 10$\ T 
{and under a certain pressure $P_{\rm c}>2$\ GPa, which is} in consistent with the experiment 
{reported in} Ref.\ \cite{Hirose}.  
{This physical picture is also consistent with the disappearance of magnetic correlations 
in the region $H\gsim 2.5$\ T, which was observed by the inelastic neutron scattering experiment 
\cite{Rossat-Mignod}. }

The heavy fermion compound CeRu$_2$Si$_2$, exhibiting metamagnetic behavior, 
is also expected to exhibit a valence transition at $P\sim 4$ GPa, offering us 
another candidate for investigating the effect of magnetic field on valence 
transition \cite{Flouquet2}. 


\section{Signature of sharp valence crossover in CeRhIn$_5$ under pressure}
\begin{figure}
\begin{center}
\includegraphics[width=0.9\linewidth]{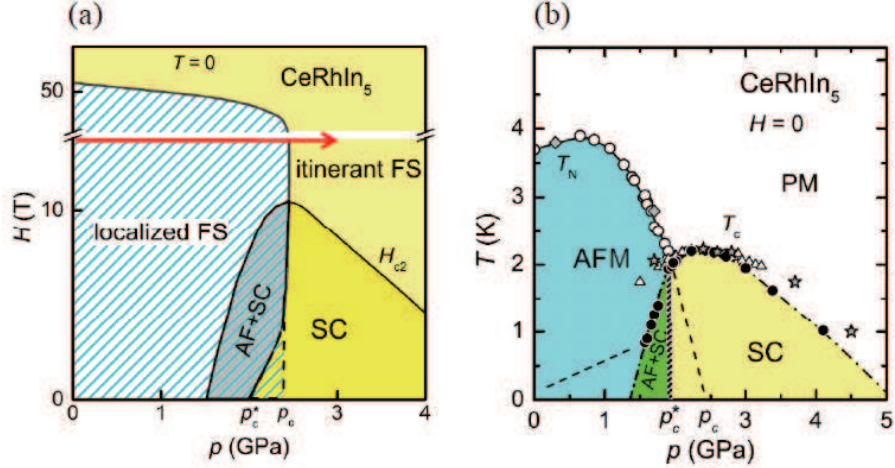}
\caption{(Color online) 
(a) Phase diagram of CeRhIn$_5$ at $T\to 0$~K in $P$-$H$ plane.~\cite{Knebel} 
(b) Phase diagram of CeRhIn$_5$ without magnetic field ($H=0$~Tesla) in $P$-$T$ plane \cite{Knebel}. 
The dashed line indicates the superconducting transition temperature reported in Ref.\ \cite{Chen}. 
}
\label{Fig:5}
\end{center}
\end{figure}

CeRhIn$_5$ is one of prototypical heavy fermion systems well studied experimentally. 
Its phase diagram is shown in Fig.\ \ref{Fig:5}: (a) in the $P$-$H$ plane at $T=0\,$K, and 
(b) in the $P$-$T$ plane at $H=0$ \cite{Knebel}. 
Other than the coexistence of superconductivity and antiferromagnetic or the proximity of 
them {across} the discontinuous phase boundary under pressure \cite{Knebel,Park}, 
the de Haas-van Alphen (dHvA) measurement performed 
along the line $H\simeq15\,$Tesla (indicated by an arrow)  in Fig.\ \ref{Fig:5}(a) revealed 
the following remarkable facts \cite{Shishido}: 
\begin{enumerate}
\item[a)]
The Fermi surfaces change at $P=P_{\rm c}$ from those expected for localized f electrons 
(giving no contribution to the Fermi volume as in LaRhIn$_5$) to those for itinerant f electrons 
contributing to the Fermi volume. 
\item[b)]
The cyclotron mass exhibits a sharp peak at 
around $P=P_{\rm c}$ even though the magnetic transition is the first order and is not associated with 
AF critical fluctuations.  
\end{enumerate}
These aspects are explained naturally as a phenomenon associated with the 
valence crossover of Ce ion on the basis of the EPAM Hamiltonian defined by 
Eq.\ (\ref{eq:1}) {\cite{WM:JPSJ,WM:JPCM2}}. 
Characteristic of our theory is that the first order 
magnetic transition is not accompanied by the localization of f electrons in the magnetic phase 
but remains itinerant with mass enhancement of quasiparticles, 
which is consistent with experiments \cite{Shishido}.  
This is in marked contrast to the local criticality theory on the so-called Kondo breakdown idea 
\cite{Si2,Si,Coleman} {in which the c-f hybridization in the AF phase is completely vanishing}. 

The relation between AF order and valence transition or sharp crossover can be understood on 
the EPAM [Eq.\ (\ref{eq:1})]  by treating it in the mean-field approximations both 
for the AF order and the slave boson which is introduced to take into account 
the strong local correlation effect between on-site f electrons \cite{WM:JPSJ,WM:JPCM2}.  
In order to simulate the above two facts observed in dHvA experiment, the dispersion of conduction 
electrons is set as   
\begin{equation}
\epsilon_{k}=-2t(\cos k_{x}a+\cos k_{y}a),
\label{dispersion}
\end{equation}
which simulates the two dimensional $\beta_2$-branch observed by the dHvA measurements in 
Ref.\ \cite{Shishido}.

\begin{figure}[t]
\begin{center}
\includegraphics[width=0.8\linewidth]{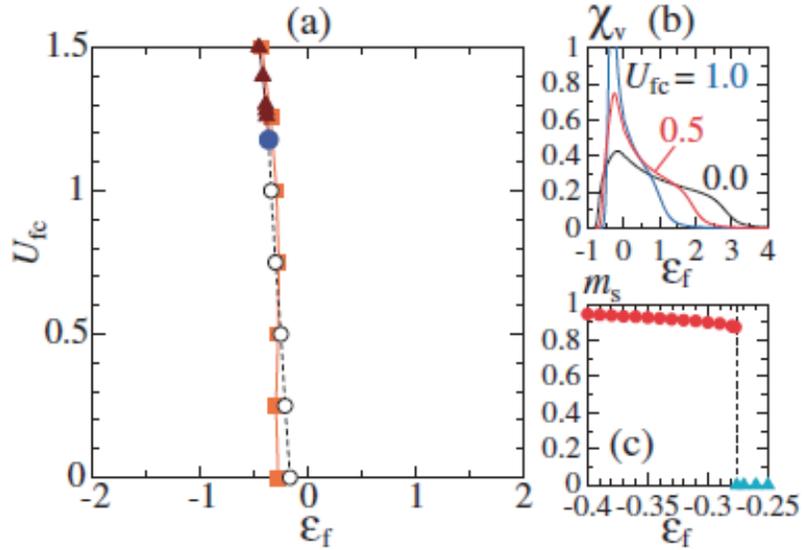}
\caption{\label{Fig:AF_Valence_1}(color) 
(a) Ground-state phase diagram in the $\varepsilon_{\rm f}$-$U_{\rm fc}$ plane   
for paramagnetic and AF states. 
The first-order valence-transition line (solid line with triangles) 
terminates at the QCEP of valence transition (filled circle). Valence crossover 
occurs at the dashed line with open circles, where $\chi_{\rm v}$ has a maximum, 
as shown in (b). 
The solid line with squares represents the boundary between the AF state 
and the paramagnetic state. 
(c) AF order parameter $m_{\rm s}$ vs $\varepsilon_{\rm f}$ 
for $U_{\rm fc}=0.5$. All results in (a)-(c) are calculated for 
$t=1$, $V=0.2$, $U=\infty$, and the electron filling ${\bar n}\equiv(n_{\rm c}+n_{\rm f})/2=0.9$ \cite{WM:JPSJ}. 
}
\end{center}
\end{figure}

Figure\ \ref{Fig:AF_Valence_1}(a) shows the phase diagram in the ground state of EPAM [Eq.\ (\ref{eq:1})], 
with the conduction electron dispersion given by Eq.\ (\ref{dispersion}), in the 
$\varepsilon_{\rm f}$-$U_{\rm fc}$ plane for the parameter set $t=1$, $V=0.2$, $U=\infty$ and the electron 
filling ${\bar n}=0.9$ \cite{WM:JPSJ}.  
A remarkable aspect is that the line of first-order valence-transition (shown by solid line with 
triangles) and that of valence crossover (shown by dashed line with open circles) almost coincides 
with the boundary between AF and paramagnetic phase (shown by solid line with squares).  
A solid circle represents the QCEP of valence transition.   
Figure\ \ref{Fig:AF_Valence_1}(b) shows the valence susceptibility 
$\chi_{\rm v}\equiv-(\partial n_{\rm f}/\partial \varepsilon_{\rm f})$ which exhibits 
clear maxima on the line of valence crossover. Namely, the AF order is cut by 
the valence transition or valence crossover \cite{Comment}. 
Figure\ \ref{Fig:AF_Valence_1}(c) shows the $\varepsilon_{\rm f}$ dependence of the AF order 
parameter  $m_{\rm s}$ exhibiting the first order transition  at $\varepsilon_{\rm f}=0.283$.  
    
\begin{figure}[t]
\begin{center}
\includegraphics[width=0.8\linewidth]{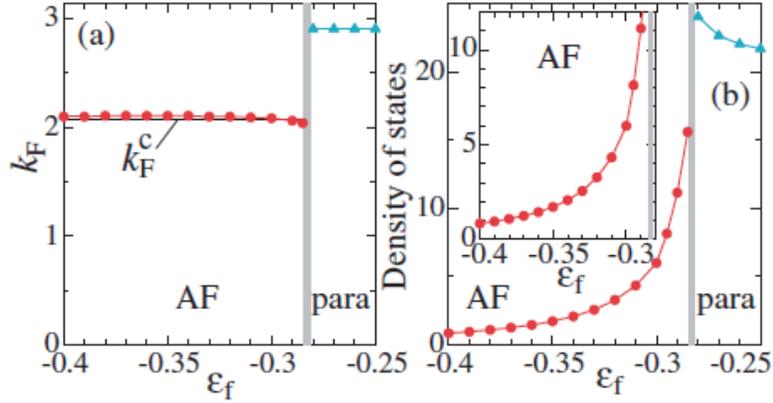}
\caption{\label{Fig:AF_Valence_2}(color) 
(a) Fermi wave number $k_{\rm F}$ vs $\varepsilon_{\rm f}$ in 
the AF state (circles) and the paramagnetic state (triangles) 
for $t=1$, $V=0.2$, $U=\infty$, $U_{\rm fc}=0.5$, and the electron filling  
${\bar n}\equiv(n_{\rm c}+n_{\rm f})/2=0.9$ under the magnetic field 
$h=0.005$.  The solid line represents $k_{\rm F}$ for the conduction band $\varepsilon_{\rm k}$ at 
$\bar{n}_{\rm c}=0.4$. 
(b) Density of states at the Fermi level $D(\mu)$ vs $\varepsilon_{\rm f}$ for the same parameters 
as (a). The inset is enlargement of the AF phase \cite{WM:JPSJ}. 
}
\end{center}
\end{figure}

Figure\ \ref{Fig:AF_Valence_2}(a) shows the change of the Fermi wave number $k_{\rm F}$ 
along the line of $k_x=k_y$ 
for the parameter set  $t=1$, $V=0.2$, $U=\infty$, $U_{\rm fc}=0.5$, and the electron filling 
${\bar n}\equiv(n_{\rm c}+n_{\rm f})/2=0.9$ {\it under the magnetic field} $h=0.005$
with the Hamiltonian  [Eq.\ (\ref{eq:1_Zeeman})]\cite{WM:JPSJ}. 
This shows that, 
associated with the onset of AF state by the first order transition, the Fermi surface 
changes discontinuously from {the} smaller size to {the} larger one as if the 
f-electrons were localized in 
the AF state or the transition to AF state were accompanied by the localization of f electrons. 
This is because the $k_{\rm F}$ in the AF state almost coincides with $k_{\rm F}^{\rm c}$ of the 
conduction electrons with the filling  $\bar{n}_{\rm c}\equiv(n_{\rm c}/2)=0.4$ 
since $\bar{n}_{\rm c}=n-(n_{\rm f}/2)=0.9-(n_{\rm f}/2)$  is reduced to  $\bar{n}_{\rm c}=0.4$ 
if the number of f electron per Ce is $n_{\rm f}=1.0$ corresponding to the ``localization'' of f-electrons 
at each Ce site. 
However, it is not the case. Indeed, the quasiparticles consist both of f- and conduction electrons there, 
and their effective mass is still enhanced in consistent with specific heat measurement 
in CeRhIn$_5$ \cite{Shishido2,Phillips}, 
showing that the Sommerfeld constant in the AF state $\gamma = 50$ mJ/moll$\cdot$K$^2$ is 
about 10 times
larger than $\gamma = 5.7$ mJ/mol$\cdot$K$^2$ in LaRhIn$_5$.  The origin of small Fermi surface is 
the effect of the band folding associated with the onset of the AF ordering 
\cite{WM:JPSJ}{, but not that of the localization of f electrons \cite{Si,Coleman,Si2}}. 

The enhancement in observed cyclotron mass, near the phase boundary of AF and paramagnetic 
states,  is reproduced theoretically, as shown in  Fig.\ \ref{Fig:AF_Valence_2}(b).  
The origin of this enhancement is the band effect of folding or unfolding of the Fermi surface associated 
with the AF transition{, which is} in consistent with the above picture of the change of the Fermi surface at 
the transition \cite{WM:JPSJ}.   

To summarize the above discussions, the characteristic aspects a) and b) of CeRhIn$_5$ obtained by 
the dHvA experiments \cite{Shishido}, listed in P.\ 11, can be understood in a unified way 
as a phenomenon associated with the valence crossover of Ce ion under the pressure and 
the magnetic field.

Other circumstantial evidence for the sharp valence crossover to be realized in CeRhIn$_5$ 
at $P=P_{\rm c}$ is summarized as follows:   
\begin{enumerate}
\item[1)]
The resistivity at $T=2.25\,$K just above $T_{\rm c}$ exhibits 
huge peak at around $P=P_{\rm c}$~\cite{Knebel,Muramatsu},  which is a 
reminiscent of the case of CeCu$_2$(Si,Ge)$_2$.  This indicates that the valence fluctuations 
are enhanced around $P=P_{\rm c}$, as discussed in Ref.\ \cite{MM}.   
\item[2)] 
The exponent of $\alpha$, in $[\rho(T)-\rho_{0}]\propto T^{\alpha}$, approaches 1 near $P=P_{\rm c}$, 
as demonstrated in Ref.\ \cite{Park}.   
This gives the signature of existence of critical valence fluctuations \cite{Holmes}.  
\item[3)] 
The so-called  Kadowaki-Wood scaling, $\sqrt{A}/m^{*}={\rm const}.$, 
holds at $P\lsim P_{\rm c}$, while both $A$ and $m^{*}$ 
grow steeply as $P$ approach $P_{\rm c}$ from the lower pressure side.   
This implies that the divergent behaviors in $A$ and $m^{*}$ are not due to the AF critical fluctuations 
which would make $A/(m^{*})^{2}$ diverge there.  
\end{enumerate}

Concluding this section, we here discuss a symptom suggesting the existence of  {the} 
QCEP of  {the} valence transition on the 
phase boundary between AF and paramagnetic state (at $P\simeq P_{\rm c}$) at higher magnetic field 
outside the region explored experimentally so far.  A crucial point is 
that the pressure dependence of the SC transition temperature $T_{\rm c}$ 
and the upper critical field $H_{{\rm c}2}$ are quite different (see Fig.\ \ref{Fig:5}).  
Namely, the former is almost flat at $P\gsim P_{\rm c}$ [Fig.\ \ref{Fig:5}(a)] 
while the latter prominently increases as the pressure approaches $P_{\rm c}$ [Fig.\ \ref{Fig:5}(b)]. 
This suggests that the SC pairing interaction is promoted 
by the magnetic field $H$.  One of such possibilities is that the QCEP of valence 
transition is located at the magnetic field $H=H^{*}$ ($>10\,$Tesla) on the phase boundary 
at $P=P_{\rm c}$  
between the AF and the paramagnetic state in the phase diagram Fig.\ \ref{Fig:5}(a).  This is reasonable, 
considering the fact that its phase boundary coincides with the valence crossover lines as shown 
in Fig.\ \ref{Fig:AF_Valence_1} \cite{WM:JPSJ}, and the SC state is stabilized in the region where 
a sharp crossover of valence occurs \cite{Onishi1,WIM,Holmes,Miyake1}.  

\begin{figure}
\begin{center}
\includegraphics[width=0.55\linewidth]{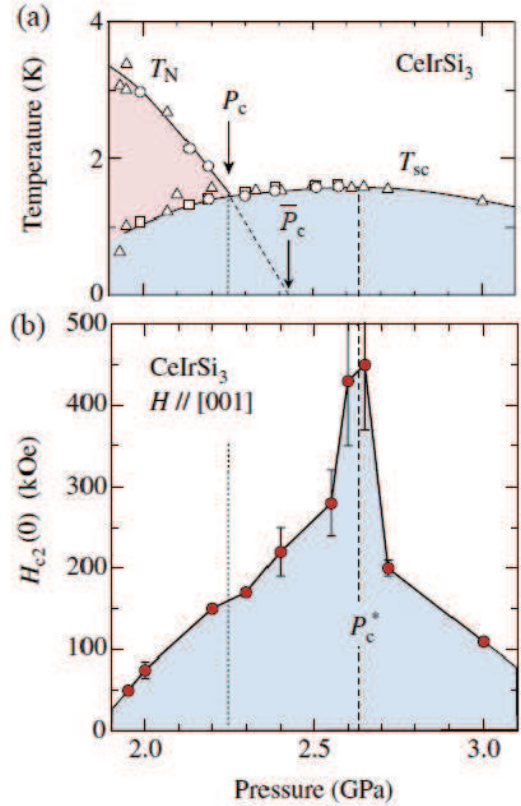}
\caption{(Color online) 
(a) $P-T$ phase diagram of CeIrSi$_3$ without magnetic field, i.e., $H=0$. 
$P_{\rm c}$ is defined as the pressure where 
the N\'{e}el temperature $T_{\rm N}$ coincides with the superconducting transition temperature 
$T_{\rm sc}$ \cite{Settai}, and ${\bar P}_{\rm c}$ is a hypothetical critical pressure where the $T_{\rm N}$ 
extrapolated into the SC phase vanishes.  
(b) Upper-critical field $H_{{\rm c}2}(T\to 0)$ as a function of $P$. $H_{{\rm c}2}(T\to 0)$ takes sharply 
enhanced maximum at $P_{\rm c}^{*}$, where the $T_{\rm sc}$ takes a broad maximum at 
$H=0$ \cite{Settai}.      
}
\label{Fig:15}
\end{center}
\end{figure}

It is also interesting to note that a similar trend can be seen in CeIrSi$_3$ \cite{Settai} and 
CeRhSi$_3$ \cite{Kimura}.  Indeed, Fig.\ \ref{Fig:15}(a) shows the $P-T$ phase diagram of 
CeIrSi$_3$ \cite{Settai}, in which the pressure 
${\bar P}_{\rm c}$, where the smooth extrapolation of the N\'{e}el temperature $T_{\rm N}$ in 
the SC phase 
vanishes, and the pressure $P_{\rm c}^{*}$, where the SC transition temperature $T_{\rm sc}$ 
takes a broad maximum, is markedly different.  Furthermore, the uppercritical field $H_{{\rm c}2}$ 
exhibits a sharp and huge enhancement around $P=P_{\rm c}^{*}$, as shown in Fig.\ \ref{Fig:15}(b).  
This implies that the pairing interaction is increased sharply as the magnetic field $H$ is increased, 
suggesting that the QCEP of valence transition is induced by the magnetic field at around 
$P=P_{\rm c}^{*}$ 
as in the case of CeRhIn$_5$ at $P=P_{\rm c}$ shown in Fig.\ \ref{Fig:5}, although the definition of 
$P=P_{\rm c}$ and $P=P_{\rm c}^{*}$ are different in these two references \cite{Settai} and \cite{Knebel}. 
We also note that  {a} similar enhancement of $H_{{\rm c}2}$ was observed 
in UGe$_2$ at $P\gsim P_{x}$ \cite{Sheikin}, 
which was explained nicely by the effect that the paring interaction is enhanced by 
the effect of field induced magnetic transition between two types of ferromagnetic states in 
UGe$_2$ \cite{WM:UGe_2}.  
   
\section{Summary}
We have presented a story how the idea of the critical valence transition or sharp valence crossover 
of heavy fermion metals has been developed, and how these phenomena are ubiquitous than first thought 
in the beginning of the present century. Although we have discussed relatively well established 
cases, there would be other potential systems which have not been well recognized so far.

\section*{Acknowledgements}
We acknowledge J. Flouquet, A. T. Holmes, D. Jaccard, H. Maebashi, O. Narikiyo, Y. Onishi, and A. Tsuruta 
for long collaborations on physics of critical valence fluctuations on which a considerable part of 
the present article is based.  Discussions and conversations with  K. Deguchi, K. Fujiwara, H. Kobayashi, 
T. C. Kobayashi, K. Kuga, Y. Matsumoto, S. Nakatsuji, and N. K. Sato on experimental data 
are also acknowledged.  
This work is supported by the Grant-in-Aid for Scientific Research (No. 24540378, No. 25400369, 
{No. 15K05177, and No. 16H01077}) 
from the Japan Society for the Promotion of Science. 
One of us (S.W.) was supported by JASRI (Proposal No. 0046 in 2012B, 2013A, 2013B, 
{2014A, 2014B, and 2015A}). 

\end{document}